\documentclass[aps,prl,preprint,nopacs,superscriptaddress]{revtex4}

\usepackage{graphicx}
\usepackage{verbatim}
\usepackage{mathrsfs}
\usepackage{amsmath,amsfonts,amssymb}
\usepackage{graphicx}

\newcommand{\bee}{\begin{equation}}
\newcommand{\ee}{\end{equation}}
\def\3{2.8in}    
\def\2{2.5in}
\def\4{3.0in}\def \beq {\begin{equation}}
\def \eeq {\end{equation}}
\usepackage[paperwidth=8.5in,paperheight=11in,centering,hmargin=2cm,vmargin=2.5cm]{geometry} 
\allowdisplaybreaks 

\newcommand{\bs}[1]{\boldsymbol{#1}}

\begin{document}

\title{Drumhead Surface States and Topological Nodal-Line Fermions in TlTaSe$_2$}

\author{Guang~Bian\footnote{These authors contributed equally to this work.}\footnote{gbian@Princeton.edu}}
\affiliation {Laboratory for Topological Quantum Matter and Spectroscopy (B7), Department of Physics, Princeton University, Princeton, New Jersey 08544, USA}

\author{Tay-Rong~Chang$^*$}
\affiliation {Department of Physics, National Tsing Hua University, Hsinchu 30013, Taiwan}
\affiliation {Laboratory for Topological Quantum Matter and Spectroscopy (B7), Department of Physics, Princeton University, Princeton, New Jersey 08544, USA}

\author{Hao~Zheng}
\affiliation {Laboratory for Topological Quantum Matter and Spectroscopy (B7), Department of Physics, Princeton University, Princeton, New Jersey 08544, USA}

\author{Saavanth Velury}
\affiliation {Department of Physics, University of California, Berkeley, California 94720, USA}

\author{Su-Yang~Xu}
\affiliation {Laboratory for Topological Quantum Matter and Spectroscopy (B7), Department of Physics, Princeton University, Princeton, New Jersey 08544, USA}

\author{Titus Neupert}
\affiliation {Laboratory for Topological Quantum Matter and Spectroscopy (B7), Department of Physics, Princeton University, Princeton, New Jersey 08544, USA}
\affiliation {Princeton Center for Theoretical Science, Princeton University, Princeton, New Jersey 08544, USA}

\author{Ching-Kai Chiu}
\affiliation {Condensed Matter Theory Center, Department of Physics, University of Maryland, College Park, Maryland 20742-4111}

\author{Shin-Ming Huang}
\affiliation {Centre for Advanced 2D Materials and Graphene Research Centre National University of Singapore, 6 Science Drive 2, Singapore 117546}
\affiliation {Department of Physics, National University of Singapore, 2 Science Drive 3, Singapore 117542}

\author{Daniel S. Sanchez}
\affiliation {Laboratory for Topological Quantum Matter and Spectroscopy (B7), Department of Physics, Princeton University, Princeton, New Jersey 08544, USA}

\author{Ilya~Belopolski}
\affiliation {Laboratory for Topological Quantum Matter and Spectroscopy (B7), Department of Physics, Princeton University, Princeton, New Jersey 08544, USA}

\author{Nasser Alidoust}
\affiliation {Laboratory for Topological Quantum Matter and Spectroscopy (B7), Department of Physics, Princeton University, Princeton, New Jersey 08544, USA}

\author{Peng-Jen Chen}
\affiliation {Department of Physics, National Taiwan University, Taipei 10617, Taiwan}
\affiliation {Nano Science and Technology Program, Taiwan International Graduate Program, Academia Sinica, Taipei 11529, Taiwan and National Taiwan University, Taipei 10617, Taiwan}
\affiliation {Institute of Physics, Academia Sinica, Taipei 11529, Taiwan}

\author{Guoqing Chang}
\affiliation {Centre for Advanced 2D Materials and Graphene Research Centre National University of Singapore, 6 Science Drive 2, Singapore 117546}
\affiliation {Department of Physics, National University of Singapore, 2 Science Drive 3, Singapore 117542}

\author{Arun Bansil}
\affiliation {Department of Physics, Northeastern University, Boston, Massachusetts 02115, USA}

\author{Horng-Tay Jeng}
\affiliation {Department of Physics, National Tsing Hua University, Hsinchu 30013, Taiwan}
 \affiliation {Institute of Physics, Academia Sinica, Taipei 11529, Taiwan}
 
\author{Hsin Lin\footnote{nilnish@gmail.com}}
\affiliation {Centre for Advanced 2D Materials and Graphene Research Centre National University of Singapore, 6 Science Drive 2, Singapore 117546}
\affiliation {Department of Physics, National University of Singapore, 2 Science Drive 3, Singapore 117542}

\author{M. Zahid Hasan\footnote{mzhasan@Princeton.edu}}
\affiliation {Laboratory for Topological Quantum Matter and Spectroscopy (B7), Department of Physics, Princeton University, Princeton, New Jersey 08544, USA}
\affiliation {Princeton Center for Complex Materials, Princeton University, Princeton, New Jersey 08544, USA}

\pacs{}

\date{\today}

\begin{abstract}
A topological nodal-line semimetal is a new state of matter with one-dimensional bulk nodal lines and two-demensional ``drumhead" surface bands. Based on first-principles calculations and our effective $ \textit{\textbf{k}} \cdot  \textit{\textbf{p}}$ model, we theoretically propose the existence of topological nodal-line fermions in the ternary transition-metal chalcogenide TlTaSe$_2$.  The noncentrosymmetric structure and strong spin-orbit coupling give rise to spinful nodal-line bulk states which are protected by a mirror reflection symmetry of this compound. This is remarkably distinguished from other proposed nodal-line semimetals such as Cu$_3$NPb(Zn) in which the nodal line exist only in the limit of vanishing spin-orbit coupling and thus is not possible in real material.  
We show that the ``drumhead" surface states in TlTaSe$_2$, which are associated with the topological nodal lines, exhibit an unconventional chiral spin texture and an exotic Lifshitz transition as a consequence of the linkage among multiple ``drumhead" surface-state pockets, suggesting a new type of 2D topological metal.
\end{abstract}

\maketitle

Recently the experimental discoveries of three-dimensional (3D) topological Dirac semimetals and Weyl semimetals \cite{Weyl, Volovik, Herring, Wilczek, Balents, Review, Wan2011, Burkov2011, HgCrSe, Thallium, Hsin_TaAs, Dai_TaAs, Su_TaAs, IOP_TaAs, Hasan_Na3Bi, Nagaosa, CdAs, Hasan_Cd3As2, Borisenko_Cd3As2, Chen_Na3Bi} have stimulated enormous research interest in topological semimetals. Topological semimetallic (TS) materials are characterized by robust bulk band crossing points and the associated topological boundary states. In the 3D TS materials, the band crossing points can be either zero dimensional (0D) discrete nodal points or one dimensional (1D) continuous nodal lines. Materials that host these exotic band structures exhibit unique properties and hold promise for applications such as topological qubits, low-power electronics and spintronics.  One prominent example of TS materials with 0D band crossing points is the Weyl semimetals. The nodal points of Weyl semimetals carry non-zero chiral charges and are connected by the Fermi arc surface states. The Weyl semimetals have been experimentally realized in transition metal monopnictides such as TaAs \cite{Hsin_TaAs, Su_TaAs, Dai_TaAs, IOP_TaAs}. The other class of topological semimetals is the nodal-line semimetal whose conduction and valence bands cross each other at closed lines instead of discrete points. The topological nodal-line semimetal is distinct in three aspects compared to the Weyl semimatal: (1) the bulk fermi surface is 1D and 0D in nodal-line semimetals and Weyl semimetals, respectively; (2) the density of states of low-energy bulk excitations is proportional to $(E-E_{f})^2$ and $|E-E_{f}|$ in nodal-line and Weyl semimetals, respectively; (3) The nodal lines are accompanied by ``drumhead"-like surface states while Weyl nodal points are connected by 1D Fermi arc surface states \cite{Burkov2011_2, Phillips, Chiu, Chen_NL, Bian, Zhang}. The unique properties of nodal-line semimetals offer a new playground for studying novel physics arising from correlations between the massless quasiparticles. For example, interaction-induced instabilities that have been broadly discussed for Weyl semimetals should be more likely occurring in nodal-line states due to the higher density of states at the Fermi energy.

Up to date, there have been several theoretical proposals for a material realization of topological nodal-line semimetals \cite{Ca3P2, CuPdN, LaN, Rappe, graphene}. All these works predict nodal-line bulk states and ``drumhead" surface states. However, the stability of nodal lines in the works requires the absence of spin-orbit coupling (SOC). With the inclusion of SOC, each nodal line is gapped due to the interaction between spin components. In real materials SOC, on the other hand, is ubiquitous, therefore it is important to study nodal-line semimetals under the condition of nonvanishing SOC. Generally, spinful nodal lines are unstable, which can be seen from a simple codimension analysis \cite{Chen_NL}.  In order to have robust nodal lines in the presence of SOC, an extra crystalline symmetry is needed to protect them. In this work, we report, based on first-principles calculations, the existence of spinful topological nodal lines in the ternary transition-metal chalcogenide TlTaSe$_2$. The nodal lines are 0.22 eV below the Fermi level and protected by a mirror reflection symmetry of the space group. The topological nodal-line state in TlTaSe$_2$ is in the class A+$R$ (p = 2) of symmetry-protected semimetals \cite{Chiu, Bian} and is conneted with spin-polarized ``drumhead" surface modes. We also demonstrate an exotic Lifshitz transition as a consequence of the linkage among multiple ``drumhead" surface-state pockets. Our results establish TlTaSe$_2$ a promising material for studying nodal-line physics.

TlTaSe$_2$, shown in Fig. 1a, crystalizes in a hexagonal lattice in which the unit cell consists of one Tl, one Ta and two Se atoms and each atom resides on a hexagonal layer. The stacking sequence of these atomic planes within the unit cell is Tl-Se-Ta-Se: B-A-B-A (A, B and C, here, refer to the three high-symmetry spots on a hexagonal lattice). In other words, the Tl layer intercalates between adjacent TaSe$_2$ layers with Tl atoms aligned with Ta atoms in the vertical direction. The structure is non-centrosymmetric and belongs to the space group $P\bar{6}m2$ (187). The lattice is reflection-symmetric with respect to both the Ta plane and the Tl plane. This reflection symmetry plays a key role in protecting the topological nodal lines, as discussed later on. Figure 1b shows the bulk and (001)-projected surface Brillouin zones where the A, H and L points are high symmetry points on the $k_z$ = $\pi$ plane, a mirror plane of the bulk Brillouin zone. The calculated band structure of TlTaSe$_2$ without and with the inclusion of SOC is shown in Fig. 1c.  Around the H point, a hole pocket derived from Ta-$5d_{xy/x^{2}-y^{2}}$ orbitals crosses an electron pocket from Tl-$6p_{x,y}$ orbitals, taking the plane parallel to the Ta atomic plane as the $x$-$y$ plane.  All these atomic orbitals are invariant under the mirror reflection $R_z$ with respect to the Tl atomic plane. A zoom-in view of band structure around H is plotted in Fig.\ 1d. In the case without SOC, the conduction and valence bands belong to different representations of the space group,  A$'$ and A$''$ for electron and hole bands, respectively. The intersection of the two bands is, therefore, protected by the crystalline symmetry, forming a spinless nodal ring on the mirror plane $k_z$ = $\pi$. Upon turning on SOC, each band splits into two spin-polarized branches since the system lacks space inversion symmetry. The spin splitting results in an accidental band touching of Ta and Tl bands at H and three band crossings. Only the band crossing 0.22 eV below the Fermi level remains gapless while the other two are gapped. A detailed analysis on the symmetry, orbital composition and spin texture of the bands around H is presented in Fig. 2a. The bands are mainly comprised of Ta-$5d_{xy/x^{2}-y^{2}}$ and Tl-$6p_{x,y}$ states which are mostly confined the Ta and Tl atomic planes, respectively. The spin of these states is primarily oriented along $z$ as indicated in Fig. 2a. At the gapless crossing point, the two branches have opposite mirror parity eigenvalues. Therefore, this band crossing is protected by the mirror symmetry, forming a pair of nodal rings on the mirror plane $k_z$ = $\pi$, one around H and the other around H$'$. To visualize the nodal rings, we plot the iso-energy contour at $E$ = $-$0.18~eV, shown in Fig. 2b. The energy is slightly off the nodal-line energy ($-$0.22 eV), which creates a tubular Fermi surface enclosing nodal lines. Indeed, we can find two rings surrounding H and H$'$ points on $k_z$ = $\pi$ plane. At the center of the bulk Brillouin zone there is a spherical Fermi surface which is the hole pocket around $\Gamma$ mainly derived from the Ta-5$d_{3z^{2}-r^{2}}$ orbitals. 

To further illustrate the mechanism of mirror-symmetry protection on the nodal rings, we develop an effective $\bs{k}\cdot \bs{p}$ description. The reflection symmetry $R_z$ that sends $z$ to $-z$ in position space acts on the spin degree of freedom like the third Pauli matrix $\sigma_3$. The crystal structure of TlTaSe$_2$ is layered in the $z$ direction, where Tl and Ta layers are alternating. A consistent choice of the action of $R_z$ is 
\begin{equation}
R_z \psi_{\mathrm{Tl}|n_x,n_y,n_z}=\psi_{\mathrm{Tl}|n_x,n_y,-n_z},
\qquad
R_z \psi_{\mathrm{Ta}|n_x,n_y,n_z}=\psi_{\mathrm{Ta}|n_x,n_y,-n_z-1},
\end{equation}
where $\psi_{\mathrm{Tl}|n_x,n_y,n_z}$ and $\psi_{\mathrm{Ta}|n_x,n_y,n_z}$ are the single-particle wave functions of an electrons in the Tl and Ta orbitals in the unit cell indexed by the integers $n_x$, $n_y$, $n_z$.
For Bloch states in the $k_z=\pi$ plane, the action of $R_z$ is local in momentum space. If $\psi_{k_x,k_y,k_z}$ is a 4-spinor in the space of spin and Tl/Ta orbital degrees of freedom, then $R_z$ has the representation
\begin{equation}
R_z\psi_{k_x,k_y,\pi}=\tau_3\otimes\sigma_3\psi_{k_x,k_y,\pi},
\end{equation}
where the Pauli matrices $\sigma_i$ act on the spin space and the Pauli matrices $\tau_i$ act on the Tl/Ta orbital space. 

We now give an effective $\bs{k}\cdot \bs{p}$ Hamiltonian to explain how the line nodes centered around the H/H$'$ points arise. We will focus on H, implying that the Hamiltonian around H$'$ follows as its time-reversal symmetric conjugate. We observe from the first principal calculations that the dispersions for the Tl and Ta orbitals are particle- and hole-like, respectively, around these high-symmetry points. 
In absence of spin-orbit coupling, the Hamiltonian for small momenta away from H is given by
\begin{equation}
H^{0}_{H}(k_x,k_y,k_z)=
\left(\frac{k_x^2+k_y^2}{2m}-\mu\right)\tau_3\otimes\sigma_0+vk_z\tau_1\otimes\sigma_0,
\label{eq: HH base}
\end{equation}
which commutes with the representation $\tau_3\otimes\sigma_3$ of $R_z$ within the mirror plane.

We will include spin-orbit coupling terms that are constants to lowest order in $\bs{k}$. Relevant mass terms are given by 
\begin{equation}
H^{(\text{SOC})}_{H}(k_x,k_y,k_z)=
\Delta_{\text{SOC}}\tau_3\otimes\sigma_3
+
M_1+
M_2,
\label{eq: HH SOC}
\end{equation}
where the mass terms $M_1$ and $M_2$ can be any matrix that anticommutes with the kinetic term $\tau_3\otimes\sigma_0$, but commutes with the mirror symmetry $\tau_3\otimes\sigma_3$. This leaves four choices of mass terms, namely $\tau_1\otimes\sigma_1$, $\tau_1\otimes\sigma_2$, $\tau_2\otimes\sigma_1$, and $\tau_2\otimes\sigma_2$. All of them will be in principle present. We have included two of them in the above Hamiltonian to reproduce the qualitative feature observed in the DFT calculations that the band gap above the nodal ring is much smaller than the band gap below the nodal ring. This means that two competing (i.e., commuting) mass terms are almost of the same magnitude. For example, we can choose $M_1=m_1 \tau_1\otimes\sigma_1$ and $M_2=m_2 \tau_2\otimes\sigma_2$ as competing mass terms. 

Let us consider $H^{(\text{0})}_{H}+H^{(\text{SOC})}_{H}$ for $k_z=0$. The bands are given by
\begin{equation}
E_H(k_x,k_y,0)=\pm\Delta_{\text{SOC}}\pm\sqrt{(m_1+m_2)^2+\left(\frac{k_x^2+k_y^2}{2m}-\mu\right)^2},
\end{equation}
with uncorrelated signs. The middle two bands are degenerate along (at most) two lines with $k_x^2+k_y^2=2m[\mu\pm\sqrt{\Delta_{\text{SOC}}^2-(m_1+m_2)^2}] $. 
These two nodal lines are protected by the opposite mirror eigenvalues of the crossing bands. 
The actual band structure of TlTaSe$_2$ is close to the parameter regime where the line node with smaller radius shrinks to zero. It disappears at the parameter point
\begin{equation}
\Delta_{\text{SOC}}^2=\mu^2+(m_1+m_2)^2.
\end{equation}

The band structure from the effective Hamiltonian is plotted in Fig. 2c, which is in remarkable agreement with the first-principles result. Considering the crystalline symmetry of TlTaSe$_2$, the nodal line at H is classified as the time-reversal breaking class A+$R$ which admits an integer topological classification for Fermi surfaces of codimension 2, i.e., lines (p = 2 in \cite{Chiu, Bian}). We note that although the entire system preserves time-reversal symmetry, the effective Hamiltonian at H lacks time-reversal symmetry, which makes the nodal lines of TlTaSe$_2$ in a different class from those reported in \cite{CuPdN, Rappe}. The nodal lines of TlTaSe$_2$ are characterized by a topological quantum number $n^+$, which is given by the difference in the number of occupied bands with $R_z$ eigenvalue +1 inside and outside the nodal line. In the case at hand, $n^+$ = +1 for the nodal line at H in the $k_z$ = $\pi$ plane as shown in Fig. 2c. We also perform a band structure calculation in which that the Ta atom is slightly moved in the vertical direction and, thus, the mirror reflection symmetry is broken. In this case all of the four branches around H are found to belong to the same $S2$ representation of the reduced space group and a gap opening is allowed at every crossing point of these branches, as shown in Fig. 2d. Therefore, the nodal line in this case is gapped by mirror-symmetry-breaking perturbations. In other words, the nodal rings in TlTaSe$_2$ are indeed under the protection of the mirror reflection symmetry. 

Next we study the evolution of the nodal lines as SOC varies. The nodal-line band structure at various SOC is plotted in Fig. 3a. As shown in the above discussion, without SOC there is a spinless nodal line around H. Once turning on SOC, each band becomes two spin branches with opposite mirror parity eigenvalues and the spinless crossing line is, consequently, split into four spinful crossing lines. However, among these four crossing lines only two are robust under the protection of mirror reflection symmetry. In this case they are the crossings between bands b1 and b4 and between b2 and b3, because the mirror parity of b2 and b4 is $+1$ and that of b1 and b3 is $-1$. So there are two nodal rings around H arising from the two symmetry-protected band crossings. The crossings between b1 and b3 and between b2 and b4 are gapped. The gap size at the topmost b1-b3 crossing is very small, but it can be evidently seen in Fig.\ 2a and explained by our effective Hamiltonian as shown in Fig.\ 2c.  As SOC increases, bands b2 and b3 are gradually pulled apart from each other and with SOC = 1 in the scale relative to the real SOC of the material the two bands barely touch. For even larger SOC, e.g. SOC = 1.2, the two bands separate and, as a result, the nodal ring associated with these two bands disappears. Therefore only one nodal ring is left when SOC is beyond the critical value 1. In Fig. 3b, we plot the energy difference between the top of band b2 and the bottom of band b3 as a function of SOC.  Varying SOC, we have three different phases of nodal lines, namely, spinless nodal line, double nodal line, and single nodal line.

In order to illustrate connection of ``drumhead" surface states with the bulk nodal lines, the surface electronic structure is constructed using a first-principles-derived tight-binding model Hamilton in a slab. The projected (001)-bulk band structure along $\bar{\Gamma}-\bar{K}-\bar{M}$ direction is plotted in Fig. 4a where the nodal line is clearly seen. The surface can be terminated with either a Se layer or a Tl layer, because the bonding between Se and Tl atoms is much weaker than that between Se and Ta atoms. The surface band structure corresponding to the two possible terminations is shown in Figs. 4b and 4c. In the case of Se termination, the surface band disperses outwards with respect to $\bar{K}$ from the nodal line and grazes outwards at the edge of the bulk band below the nodal line and merges into the bulk band region, as marked by the arrows in Fig. 4b. This can be also seen in the iso-energy band contour at E = $-0.25$ eV (slightly below the energy of the nodal line, -0.22 eV), shown in Fig. 4d. The ``drumhead" surface states overlap with the bulk nodal rings and there is no other surface state at this energy. On the other hand, the surface band structure on the Tl-terminated surface is dramatically different. Along $\bar{\Gamma}-\bar{K}$ direction, starting from the nodal point, the surface band grazes outwards at the edge of the upper bulk Dirac cone and merges into the bulk band. Along $\bar{K}-\bar{M}$ direction, the ``drumhead" surface band $SS1$ disperses into the bulk band gap. Within the gap there exists a second surface band $SS2$ which joins the ``drumhead" band at $\bar{M}$ forming a Kramers pair. The spin polarization of the two surface bands is shown in Fig. 4e. The two surface bands possess opposite spin polarizations. The overall surface band structure and spin texture resemble those of the Dirac surface states of topological insulators \cite{RMP, Zhang_RMP}. The ``drumhead" surface band is without spin degenracy and it arises from the band inversion of two spinful bulk bands (b1 and b4). Figure 4f shows the iso-energy band contours of Tl-terminated surface at three different energies (indicated by dotted lines in Fig. 4e). At $E$ = $-$0.25 eV, slightly below the energy of the nodal rings, two surface bands form a human-eye-shaped contour, lying in between two nodal-ring bulk pockets at $\bar{K}$ and $\bar{K'}$. The spin texture is unique in the sense that as moving along either surface band clockwisely, the spin orientation rotate in counterclockwise direction. The spin orientation is not always in the tangential direction of the Fermi surface contour, which is distinct from the spin-momentum-locked texture of Dirac surface states in conventional topological insulators \cite{RMP, Zhang_RMP}. At the energy of the nodal rings ($E$ = $-$0.22 eV), the eye-shaped contour connects to the nodal rings and, as energy moves further up, the corners of the eyes open and the ``drumhead" surface band $SS1$  transforms into a giant closed contour surrounding  $\bar{\Gamma}$. This topological change in the band contour is known as Lifshitz transition in the electronic structure. The Lifshitz transition discussed here is special in two aspects: (1) the transition and the associate saddle-point singularity happen in spin-polarized 2D surface bands; (2) the transition relies on a linkage of multiple small pockets around $\bar{M}$ to form a giant pocket surrounding  $\bar{\Gamma}$. This exotic linkage of 2D surface bands give rise to a divergence in DOS, like the case in the van Hove singularity in topological crystalline insulators and high-$T_c$ superconductors \cite{Vidya, King, Hirsch, Chang}. Shifting chemical potential to the Lifshitz transition energy in TlTaSe$_2$ by means of chemical doping or electrical gating can potentially trigger interaction-induced instabilities such as unconventional superconductivity \cite{King} or spin/charge density waves at the surface.

In summary, topological nodal-line semimetals is a distinct class of topological materials beyond topological insulators and Weyl semimetals. In this work we propose theoretically that TlTaSe$_2$, a ternary transition-metal chalcogenide, is a promising candidate for material realization of topological nodal-line semimetals. Unlike previous proposed materials, the nodal lines in TlTaSe$_2$ is robust even with the inclusion of SOC as long as the mirror reflection symmetry with respect to the Ta atomic plane is not broken. The nodal line is 0.22 eV below the Fermi level and, thus, accessible to the conventional ARPES measurements. Through systematic surface simulations, we show the unique spinful ``drumhead" surface states on Se-terminated surface and, even more interestingly, find that Tl-terminated surface of TlTaSe$_2$ features a Lifshitz transition as a consequence of the exotic linkage of multiple ``drumhead" surface-state pockets. In light of these novel properties of the electronic band structure of TlTaSe$_2$, we establish an ideal material platform for studying unique nodal physics and electronic correlation effect of topological nodal-line semimetals.

\section{Methods}
 
 We computed electronic structures using the norm-conserving pseudopotentials as implemented in the OpenMX package within the generalized gradient approximation (GGA) schemes \cite{Perdew, Ozaki}. Experimental lattice constants were used \cite{Eppinga}. A 12 $\times$ 12 $\times$ 4 Monkhorst-Pack k-point mesh was used in the computations. The SOC effects are included self-consistently \cite{Theurich}. For each Tl atom, three, three, three, and two optimized radial functions were allocated for the $s$, $p$, $d$, and $f$ orbitals ($s3p3d3f2$), respectively, with a cutoff radius of 8 Bohr. For each Ta atom, $d3p2d2f1$ was adopted with a cutoff radius of 7 Bohr. For each Se atom, $d3p2d2f1$ was adopted with a cutoff radius of 7 Bohr. A regular mesh of 300 Ry in real space was used for the numerical integrations and for the solution of the Poisson equation. To calculate the surface electronic structures, we constructed first-principles tight-binding model Hamilton. The tight-binding model matrix elements are calculated by projecting onto the Wannier orbitals \cite{Weng}. We use Tl $p$, Ta $s$ and $d$, and Se $p$ orbitals were constructed without performing the procedure for maximizing localization.

\section{Acknowledgements}
T.R.C. acknowledges visiting scientist support from Princeton University. We thank Chuang-Han Hsu for technical assistance in the theoretical calculations. We thank Chen Fang and Andreas P. Schnyder for discussions.

\newpage

\begin{figure}
\centering
\includegraphics[width=16cm]{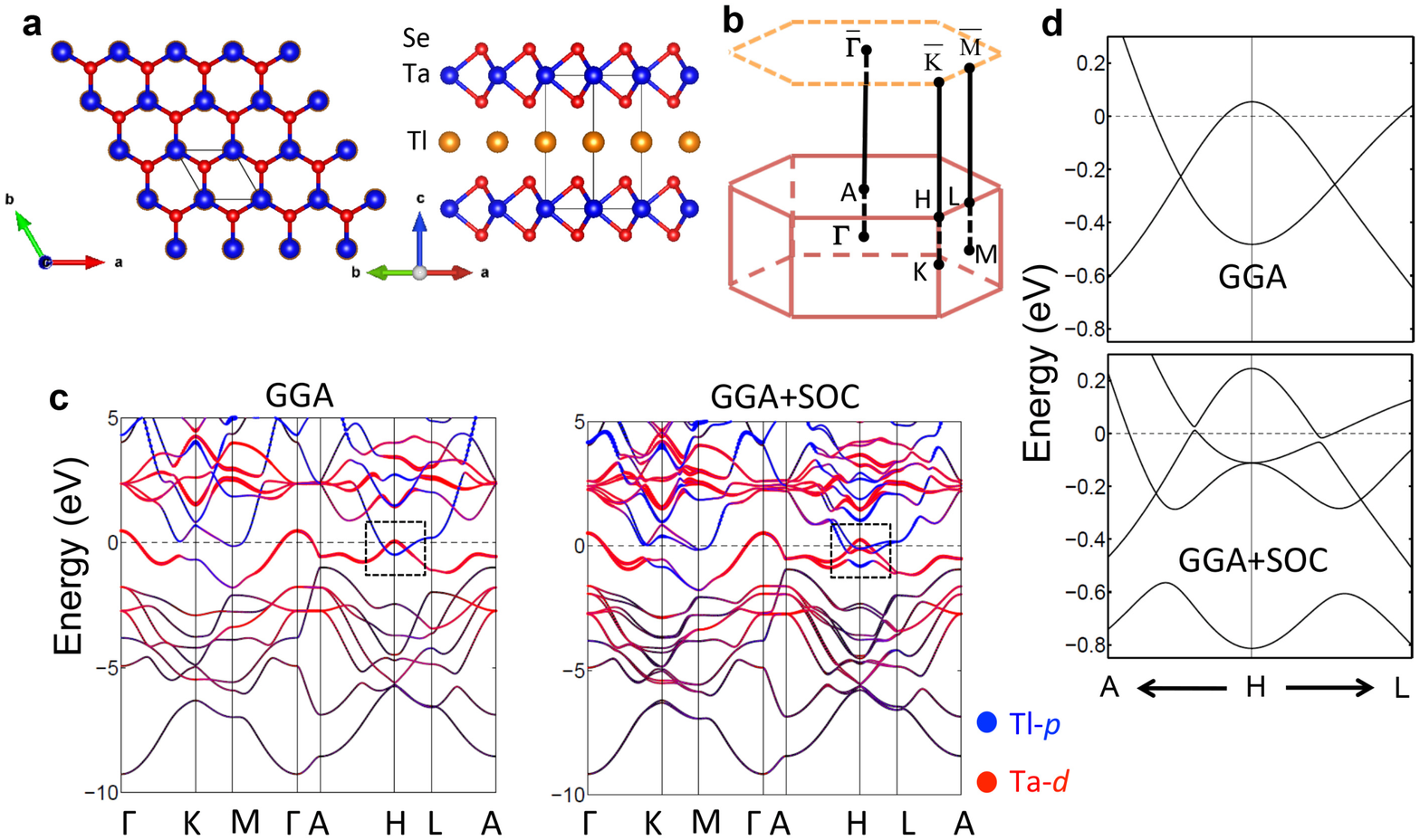}
\caption{\textbf{Crystal lattice and bulk band structure of TlTaSe$_2$.} (\textbf{a}) Lattice structure of TlTaSe$_2$: top view (left) and side view (right).  (\textbf{b}) Bulk and (001)-projected surface Brillouin zones. (\textbf{c})  Calculated bulk band structure of TlTaSe$_2$ without (left) and with (right) the inclusion of spin-orbit coupling. The color shows the atomic orbital decomposition.(\textbf{d}) Zoom-in band structure around H point as marked in \textbf{c}. } 
\end{figure}

\newpage

\begin{figure}
\centering
\includegraphics[width=16cm]{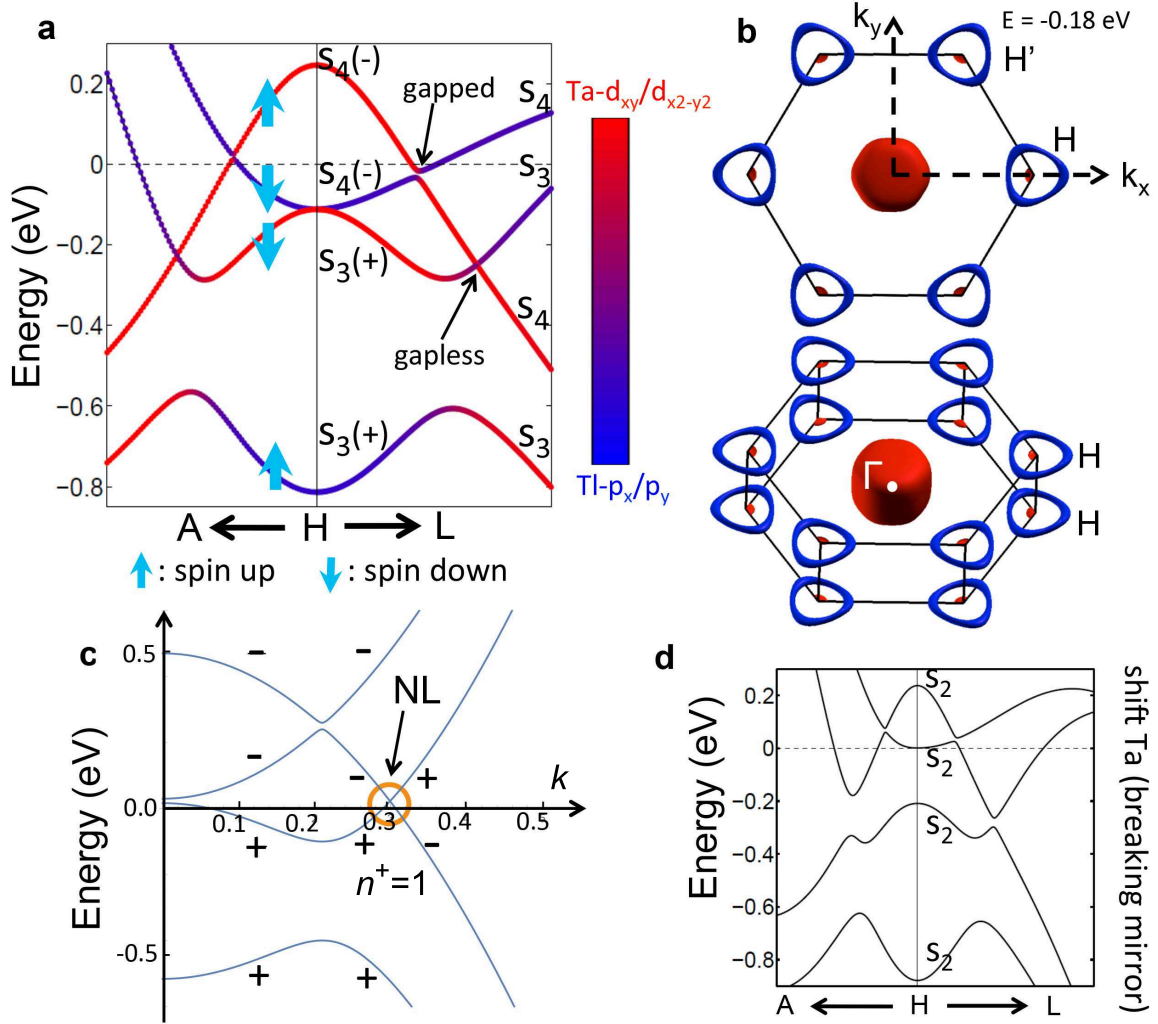}
\caption{\textbf{Symmetry, orbital composition and spin texture of nodal-line Fermions in TlTaSe$_2$.} (\textbf{a}) Orbital composition, space-group representation and spin polarization of nodal-line bands in TlTaSe$_2$. The mirror parity of each band is given in parentheses. (\textbf{b}) Iso-energy bulk-band contour at E = -0.18 eV of TlTaSe$_2$ bulk bands. (\textbf{c}) Band structure of the effective $\bs{k}\cdot\bs{p}$ model given by Eqs.~\eqref{eq: HH base} and~\eqref{eq: HH SOC} that approximates the low energy bands around the $H$ point for parameters $m_1=0.7$, $m_2=0.8$, $\mu=2.2$, $\Delta_{\mathrm{SOC}}=2.5$, with $R_z$ reflection eigenvalues of the bands and reflection-protected crossings (orange circles) that corresponds to the nodal line. All bands are nondegenerate and the momentum $k=\sqrt{k_x^2+k_y^2}$, $k_z=0$ is measured relative to the high-symmetry point $H$. (\textbf{d}) Calculated bulk band structure of TlTaSe$_2$ with the Ta atom  shifted slightly away from the equilibrium position in the unit cell. The shift breaks the mirror symmetry of the system and reduces the symmetry of the nodal-line states.} 
\end{figure}

\newpage

\begin{figure}
\centering
\includegraphics[width=16cm]{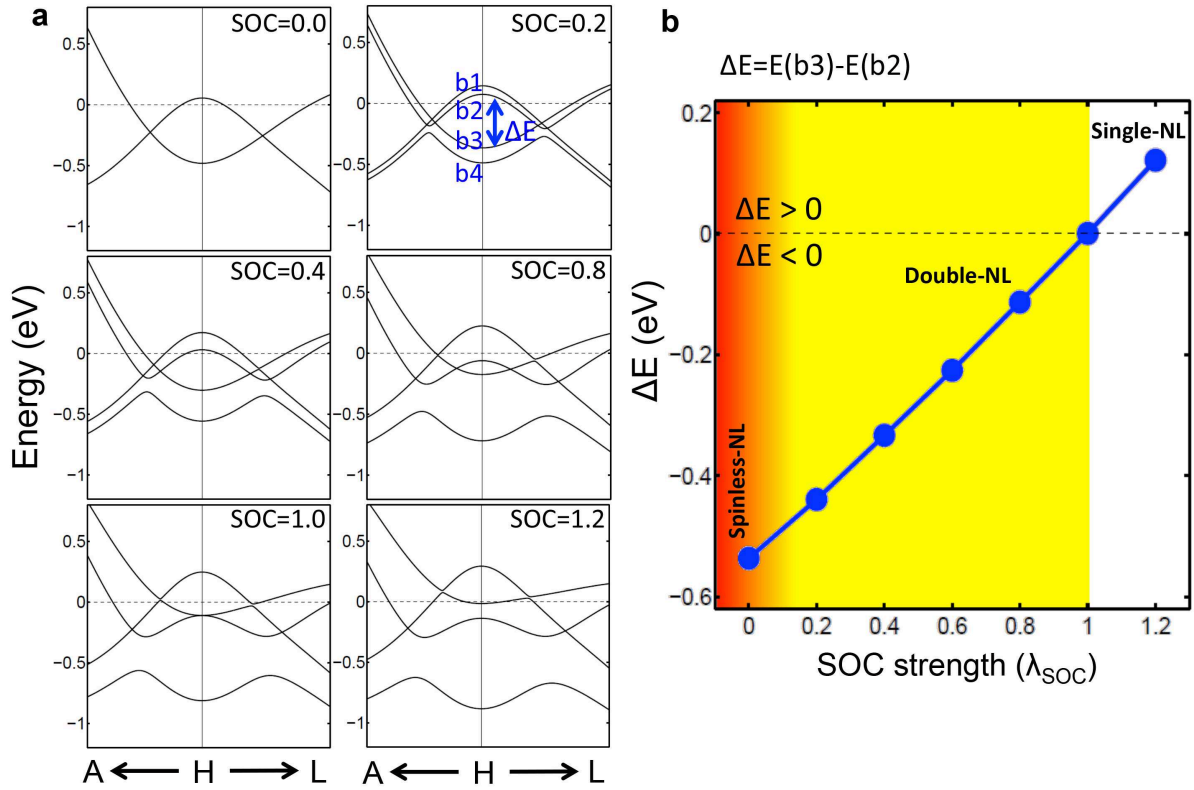}
\caption{\textbf{Evolution of the bulk band structure of TlTaSe$_2$ as the spin-orbit coupling increases.} (\textbf{a}) The nodal-line band structure of TlTaSe$_2$ with various SOC. (\textbf{b})  The energy difference between the two bands b2 and b3 as a function of SOC, showing nodal-line phases at different SOC.}
\end{figure}

\newpage

\begin{figure}
\centering
\includegraphics[width=16cm]{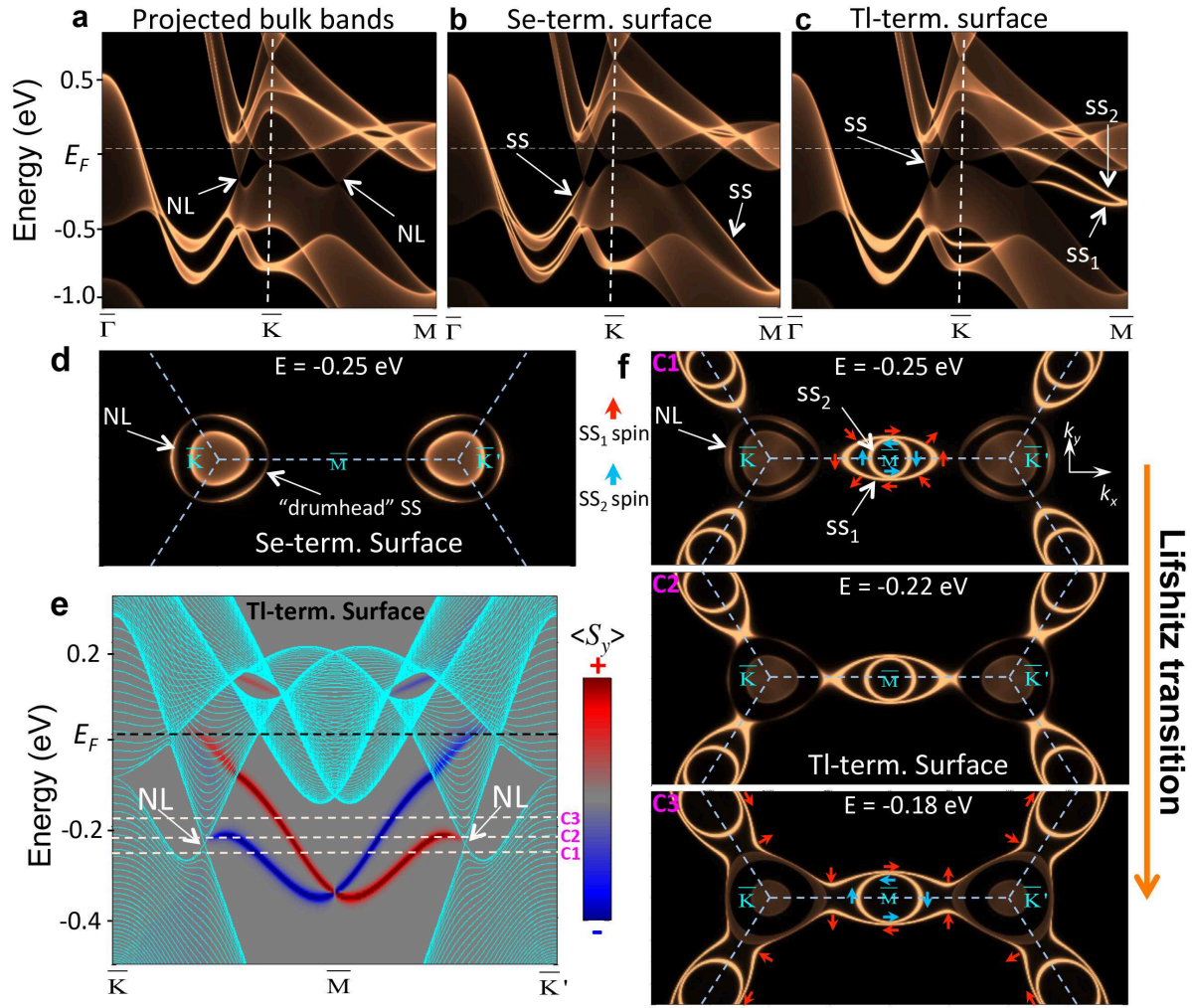}
\caption{\textbf{Nodal-line fermions and ``drumhead" surface states of TlTaSe$_2$.} (\textbf{a})  (001)-projected bulk bands of TlTaSe$_2$. (\textbf{b}) Bulk and surface band structure of Se-terminated surface of TlTaSe$_2$. The surface bands are marked by the arrows. (\textbf{c}) Same as \textbf{b} but for Tl-terminated surface of TlTaSe$_2$. (\textbf{d}) Iso-energy band contour at E = -0.25 eV of Se-terminated surface. The ``drumhead" surface states overlap with the bulk nodal rings. (\textbf{e}) The spin polarization of surface bands of Tl-terminated surface.  (\textbf{f})  Iso-energy band contour at E = -0.25 eV (top), -0.22 eV (middle), and -0.18 eV (bottom) of Tl-terminated surface. The arrows indicate the spin texture of the surface bands.}
\end{figure}

\end{document}